\documentclass[sigconf]{acmart}
\AtBeginDocument{%
  \providecommand\BibTeX{{%
    \normalfont B\kern-0.5em{\scshape i\kern-0.25em b}\kern-0.8em\TeX}}}

\setcopyright{none}
\copyrightyear{2024}
\acmYear{2024}
\acmDOI{XXXXXXX.XXXXXXX}

\acmConference[HRI'24]{March 11--15, 2024}{Designing and Intro to HRI Course Workshop}{Boulder, CO}

\usepackage{xcolor}

\begin{document}

\title{HRI Curriculum for a Liberal Arts Education}


\author{Jason R. Wilson}
\affiliation{%
  \institution{Franklin \& Marshall College}
  \city{Lancaster}
  \state{Pennsylvania}
  \country{USA}}
\email{jrw@fandm.edu}

\author{Emily Jensen}
\affiliation{%
  \institution{University of Colorado Boulder}
  \city{Boulder}
  \state{Colorado}
  \country{USA}}
  \email{emily.jensen@colorado.edu}

\renewcommand{\shortauthors}{Wilson and Jensen}

\begin{abstract}
  In this paper, we discuss the opportunities and challenges of teaching a human-robot interaction course at an undergraduate liberal arts college. We provide a sample syllabus adapted from a previous version of a course.
\end{abstract}


\maketitle

\section{Introduction}


Human-robot interaction (HRI) ought to be studied as an interdisciplinary field that examines both the technology that makes it possible and the diverse and complex implications derived from interacting with people.  
HRI as a field of research brings together engineers, psychologists, designers, anthropologists, sociologists, and philosophers \cite{bartneck2020human}, and an HRI course should examine many of these perspectives and ideally is accessible to students from each of these endeavors.
An interdisciplinary and humanistic HRI study is an ideal course for a liberal arts college.
A liberal arts education strives to empower students to 
connect information for disparate perspectives,
develop capacity for ethical and moral judgments, and
build knowledge of cultures to apply skill in complex intercultural contexts \cite{king2007liberal}.

In this paper, we present key elements of a HRI course that we are developing.  It is partially based on a course taught in the Spring of 2022, examines lessons learned, and integrates new perspectives.  There is a focus on integrating different disciplines, promoting broad perspectives, and offering learning opportunities to cultivate skills in technology and the humanities.
The next section provides relevant excerpts from the syllabus we are developing.  We then conclude with a discussion of the syllabus and connect it to the goals of a liberal arts education.

\section{Sample Syllabus}
The syllabus included here is written for a one-semester course that meets twice per week for 80 minutes. Students are expected to have taken the equivalent of a CS1 course, which introduces basic programming concepts.

\subsection{Course Description}
This course will focus on the emerging field of Human-Robot Interaction (HRI). This multidisciplinary research area draws from robotics, artificial intelligence, human-computer interaction, psychology, philosophy, and more. The main goal of HRI is to enable robots to successfully interact with humans. As robots increasingly make their way into functional roles in everyday human environments (like homes, schools, and hospitals), we need them to be able to interact with everyday people. In this course, we will learn how robots use computational models to have natural and intuitive social interactions with humans.

\subsection{Learning Outcomes}
By the end of this course, you should be able to:
\begin{itemize}
    \item Apply concepts from different disciplines in the study of HRI. Understand how interdisciplinary perspectives may compliment or contradict each other.
    \item Recognize relevant facets in designing a human-robot interaction and use relevant design tools.
    \item Identify relevant HRI topics in the media and everyday experiences. Connect these experiences with current research in HRI and discuss social and ethical implications.
    \item Apply the appropriate HRI technologies (e.g., artificial intelligence, sensors, interfaces) for a given problem. Understand the strengths, limitations, and potential biases of each.
    \item Identify the social and ethical implications resulting from the robot, how it interacts, and the technologies applied. Understand how social interaction contexts vary between cultures. 
    \item Communicate technical research to a diverse audience.
\end{itemize}

\subsection{Materials and Resources}

The book \textit{\href{https://www.human-robot-interaction.org/}{Human-Robot Interaction - An Introduction}} is available for free online.  Many chapters from this book will be assigned, especially in the first half of the course. Additional reading may also be assigned.  For each additional reading, either a link or a file will be posted on Canvas.  Students do not need to buy any reading material for this course.

For the ``HRI in the wild'' assignments, students will have the option to use news articles or other current media to highlight relevant class topics. Some possible (free) outlets to find relevant content are the \href{https://apnews.com/}{Associated Press}, \href{https://www.wired.com/}{Wired}, YouTube news segments, and webcomics (here's an \href{https://xkcd.com/1613/}{example from XKCD}). 

Students will also be regularly assigned media (e.g., videos, literature) to analyze. Most of these will be posted on Canvas. We will make every effort to use content that is freely available or provide alternative options that reduce costs.

There will be a variety of software used in this course.  This may include Visual Studio and Choregraphe. 
Most of the software should be freely available.  In most cases, the software will also be already installed on computers in the classroom or labs. Not all students will have a significant coding background, so we will aim to make the technical components as accessible as possible for everyone.

\subsection{Assignments}
There are five types of assignments in this course, each intending to assess your mastery of class material in a different way.

\textit{Reflections}. Most weeks, you will write a reflection on the material of the week.  This material may include assigned readings, videos, or movies.  To guide and assist you, we will provide a prompt to consider in the reflection.  You are invited to consider how the topic of the prompt relates to the assigned material. Each reflection should be roughly 500 words. Additionally, reflections are expected to address the given prompt and to demonstrate some level of understanding of the assigned material for the week.

\textit{Class participation and activities}. Class time will regularly be spent in discussions and completing various exercises.  The discussions and exercises often will then directly lead to the projects. Coming to class having read the readings, watched the assigned videos, etc. will be required to participate in the discussion and activities.   Each week, an in-class activity will have an associated assignment on Canvas for you to submit any material created during class.  We will also use this assignment to provide feedback on your participation for that week.

\textit{HRI in the wild}. Robots are becoming more prevalent in daily life. Each week, you will submit a short write-up discussing how you noticed course topics in the news and media. You may occasionally include a description of a personal experience with documentation of the event. Write-ups should be 100-200 words and demonstrate a clear connection between the event and a relevant course topic.

\textit{Mini projects}. There will be cumulative assignments throughout the course that correspond to the main themes in the course.  These small projects will be focused a particular problem or use case for HRI, specifics of which will be presented in the few couple of weeks of the course.
\begin{itemize}
    \item Perspectives of robots
    \item Interaction design
    \item Program a robot interaction
    \item AI for HRI
    \item Analyze and present an HRI study.
\end{itemize}

\textit{Final project}.  The final project is a culmination of the work done in the mini projects and will allow you to examine  the design and implementation of HRI work in a real-life example.
The final project will be done in a group.  The project will consist of a series of smaller assignments that will be due over the final 4-5 weeks of the course.  The assignments will be a combination of group and individual assignments.

\subsection{Schedule}
The course will be organized around the following themes. We include guiding questions that inform the content for each theme.

\textit{Theme 1: Introduction to HRI}. Understand fundamental concepts of HRI. What is a robot? What is a social robot? What modalities can a robot use when interacting with a human? Mini project: Perspectives of robots.

\textit{Theme 2: Design}. Apply design principles to HRI. What are some of the key considerations in designing a robot that will interact with a human? ... that will participate in society? Mini project: Interaction design.

\textit{Theme 3: Technology}. Implement a technical solution for a human-robot interaction. What programs control a robot? How does a robot automatically respond to human input? Mini project: Program a robot interaction.

\textit{Theme 4: Artificial Intelligence}. Design and implement the appropriate use of artificial intelligence (AI) to enable or enhance a human-robot interaction. What is AI and how can it help?  What are the limitations of this technology? What are limitations and challenges of artificial intelligence for HRI? Mini project: AI for HRI

\textit{Theme 5: Topics in HRI}. Overview of additional topics, based on class interests. For example, how may a robot use theory of mind to understand a user? Moral implications of the trolley problem for robots. Mini project: Analyze and present a recent HRI study.

\textit{Theme 6: HRI Studies}. Design and execute a HRI experiment. What types of questions can we try to answer with a HRI experiment? How do we design experiments? How do we measure and analyze the results?

\subsection{Policies}
\paragraph{Participation and Inclusivity}
Open discussion is a critical component of the learning experience in this course.  You will be expected to be prepared to participate in the discussion.  To facilitate a productive discussion, it is vital that all participants feel that their contributions are valued and recognized by others.  This can be facilitated if we all treat each other with respect. We recognize that the best science and learning happens when we integrate the expertise and perspectives from everyone. This class is stronger because we include students from a variety of programs on campus. 

\paragraph{Using External Resources}
Please refer to official college policies on academic misconduct. When it comes to using generative AI tools like ChatGPT, we find ourselves in a gray area regarding cheating. These tools are powerful, and we think it is important that you learn to use them effectively. In this course, we view using generative AI like we view working with a classmate on an assignment; while you can collaborate, the final submitted assignment should be your own. You can ``discuss'' questions you have on an assignment to help clarify difficult topics or dispel misunderstandings. However, the final product (code, text, presentation slides) should be created by you. If you use one of these tools, include proper attribution in your submitted assignment and include the prompts you used (or a link to your chat history). This allows us to learn from each other on how to use generative AI tools effectively and understand how much of the assignment came from your own thoughts and effort.

\section{Discussion}

\subsection{Assignments}
While this course covers technical content, many assignments involve a writing component. This encourages students to clearly articulate their ideas and verify their understanding of complex concepts. These assignments also make the course accessible to students without a significant coding background, which is a key difference from engineering-focused HRI courses.

\textit{Reflections}. During the second week of the course when it was previously offered, students were assigned a couple of readings
(i.e., \cite{goodrich2008human} and chapter 4 (Design) of \cite{bartneck2020human}) and to watch at least one movie from a curated list of movies involving robots (e.g., \textit{Robot \& Frank}, \textit{Big Hero 6}, \textit{Astro Boy}, \textit{The Iron Giant)}.  The students are then given the following prompt:
\begin{quote}
    How does culture influence the design of a robot?  Have you interacted with a robot before?  If so, what was the cultural context?  Or perhaps you have seen robots in movies or other media.  Are there cultural influences in how the robot is depicted?
\end{quote}
\noindent
Students brought a lot of interesting perspectives in response to this prompt, including a discussion on gendering robots
and the intricacies of pronouns and honorifics in Vietnamese
important for the deployment of robots,
particularly in eldercare, to show proper respect.
It is clear that this prompt supported the goals of liberal arts education by exercising the students skills to consider
disparate perspectives,
ethical implications,
and cultural contexts

\textit{Class Activities}. These submissions are meant to be an informal check-in about students' experience in the course and provide feedback to the instructors about topics that need more clarification. Class activities act as exit tickets, which involve one or two questions that can quickly be answered at the end of a class session. In future iterations of the class, we will consider removing formal grades for class activities; this provides students a private method for communicating with the instructor and can promote equitable engagement in the classroom \cite{fowler_exit_ticket}. An example of an exit ticket assessing a course learning outcome might be a free-response question (e.g., \textit{What is one limitation of using deep learning models for emotion detection?}) and an accompanying likert-style rating their confidence in their answer. Another example is asking about lingering questions on the current project. 

\textit{HRI in the wild}.
This assignment will be a new addition to the curriculum in the upcoming iteration of this course.  It is meant to explicitly tie the concepts learned in class to current events happening around us in the world. As a liberal arts institution, one of our goals as educators is to prepare students to take perspectives of others and understand how events happening now can impact the future. Some possible examples of current events (at the time of this submission) that students could write about include ethical implications of designing robots to replace hospitality jobs \cite{Yamat_2024} or the student's experience interacting with a floor cleaning robot at the local supermarket.

\subsection{Projects}
The mini projects and the final project are intended to focus around a particular problem or use case, particularly issues relating to students or campus life.  This can include questions like how can a robot support campus tours, events for admitted students, or students getting acclimated to campus.  There are also academic use cases, such as learning a foreign language.
In the discussion of the projects that follows, we propose a use case that was briefly explored when the course was last taught: how can a robot assist international students?

\textit{Perspectives of Robots}.
In this project, students are expected to collect a broad set of ideas related to robots, with the intention of opening their mind to new possibilities.  One important part of this would be to explore the different cultural perspectives on robots and the various roles robots may have in society. We can learn about different perspectives through media, news, conversations with other people, and literature. The format of the submitted project is up to the student; some examples are writing a synthesis essay with attached media as examples or creating an animated video. The goal of this project is for students to connect many different perspectives of what robots are/should be like in a way that makes sense to them.

\textit{Interaction design}.
This project focuses on the prospective users of the robot and explores how a robot may address the needs or desires of a set of users.  
Based on this understanding, students are tasked with proposing an example of how an interaction with the robot would play out in real life.
One approach to doing this is having an in-class activity followed by a portion that students do on their own outside of class.
Considering the focus problem of aiding international students, the in-class activity may first explore the types of issues facing international students.  A similar exercise was done when this course was last offered.  With prior permission of the international students in the class, they discussed in small groups some of the challenges they encounter.  Focusing on one challenge, the groups then created a storyboard proposing an example of how an interaction with a robot may alleviate that challenge.  The international students in the group are able to act as co-designers \cite{steen_codesign}, offering their expertise and critical perspective.
Previously, one group designed a robot that would answer students questions about local customs, from where to buy toilet paper to how much to tip at a restaurant.
After class, students would interview two other international students, presenting them first with the storyboard of the proposed interaction, and then inquiring about the strengths and weaknesses of the design.
To reduce risk of racism, xenophobia, or inappropriate assumptions about a student's nationality, students would not be asked to find international students to interview, and instead we would recruit a set of students in advance.  The international students would be compensated for their participation.

This design project works towards many of the goals of a liberal arts education.
Students will be explicitly instructed to consider potential moral or ethical implications of their design.  For example, the students many need to consider how does the robot know if a student is an international student.
With roughly 17\% of the students at Franklin \& Marshall College being international, any project designing for students will be in a multicultural context.  The example project focusing on international students emphasizes this, but similarly designing a robot to assist in campus tours would also need to consider different cultures.

\textit{Program an Interaction}.
Students will use the Choregraphe graphical software system to program an interaction for the NAO robot.
Choregraphe allows students to program portions of the interaction without a robot present, giving them flexibility on when/where to work on the project. The software does not require coding, making it accessible to students from different backgrounds. An example project might be having the robot watch the user complete a task and make a celebratory gesture when the user successfully completes the task.

In the last iteration of the course, students designed a robot interaction for the common room in campus housing.  One student had the robot present a large menu of options, from information about F\&M to entertainment. The users' favorite element was the robot jokes, such as:  \textit{What's a robot's favorite music?}  Heavy metal.  \textit{How does a robot each salsa?}  With microchips. By giving students space and the opportunity to participate in their own way, we found they created a wide variety of creative projects that simultaneously aligned with their interests and the project goals.

\textit{AI for HRI}.
This project is a new addition to the curriculum and will be adjusted in the future given feedback from students. The goal is to introduce students to off-the-shelf AI technologies that can be used to create nuanced interactions. 
Given the rapid growth of capabilities provided by large language models (LLMs) and students' likely exposure to using it, this project explores how LLMs may provide new capabilities to improve the interaction.  On the other hand, there are number of challenges and risks inherent in using a LLM.

Students will be tasked with enhancing an interaction through a creative application of a LLM.  To get the students started, they will be provided with an integration between a LLM and a social robot.  This integration gives the robot basic spoken chat capabilities.
The students may then explore
one or more of the following challenges:
\begin{itemize}
    \item Filtering inputs and/or outputs to ensure the robot interacts in an ethical and moral manner.
    \item Initializing the LLM with prompts to guide the interaction.
    \item Translate output into gestures to accompany the spoken text.
    \item Integrate more input modalities (e.g., emotion recognition).
    \item Improve roundtrip time to provide a more fluid interaction.
    \item Detect the user's spoken language and adjust the conversation accordingly.
    \item Adapt speech to use similar mannerisms and style as the user (similar to code switching \cite{gardner2009code}).
    \item Reference relevant parts of a previous conversation.
    \item Find other creative LLM uses to enhance the interaction.
\end{itemize}
Students will include in the project submission a report outlining the design decisions and justifications for those decisions.

\textit{Topics in HRI}.
For this project, students will find and present recent research.
The selected paper does not need to be directly related to the focus problem, but students are expected to connect the focus problem to the research in some way. This project challenges students to explain technical material to a diverse audience and critically think about the methods used, the problem the authors are trying to solve, and possible ethical impacts of the findings. In addition to practicing their technical reading and communication skills, students will develop an understanding of what effective and not-so-effective scientific writing looks like. 

\textit{Final Project}.
For the final project, students will work in groups to design and possibly execute an experiment related to the focus problem.  Ideally, students will be able to use the products of their mini projects in support of this experiment.  Since students did not work in groups on the mini projects, they have an opportunity to mix and match their products and thus are able to leverage each students strengths and unique perspectives.

Designing an experiment and implementing the robot controls for a complex interaction would be far too ambitious for a class project, and thus students will be invited to design the experiment around a specific research question that may be part of a larger interaction.  For example, do international students prefer a spoken, graphical, tangible interface (e.g., the contact sensors on the NAO)? Students will report their findings in a technical report, using the concepts they learned from the Topics in HRI project.

\subsection{Pedagogical Strategies}
The syllabus presented above takes advantage of several pedagogical strategies meant to maximize student engagement. The first is active learning, a student-centered method where knowledge is actively constructed during class instead of passive lectures \cite{arthurs_active_2017}. We integrate this into the syllabus by having students do much of the reading outside of class and focus on discussion and projects during class. Another pedagogical strategy is using formative feedback, an informal method of feedback that is non-evaluative and meant to guide and motivate students during their learning journey \cite{shute_focus_2008}. In this course, formative feedback is given through the class activities and on the mini-projects that will be integrated into the final project. Finally, we use elements from Universal Design for Learning by encouraging students to engage with course material in different ways and giving them options for how to communicate their understanding \cite{rose2000universal}. To make the learning expectations clear to students, each assignment will come with a concrete rubric that describes the key elements their submission should include.

\subsection{Liberal Arts Education}
We selected the content of this course to compliment a liberal arts curriculum. Many assignments are designed to incorporate technical knowledge with a reflection on the different cultural and ethical perspectives that influence the course content. In particular, we are continuing to diversify the videos, literature, and other media consumed in class to include non-western perspectives. Assignments such as reflections and HRI in the wild explicitly ask students to incorporate their world view into their responses. Working in groups for the final project allows students to question their assumptions and build off the strengths of other students who bring perspectives from other disciplines. Finally, we include what may be considered an unusual amount of writing in this course. This is meant to help students develop effective communication skills and combat the stereotype that scientists and engineers are poor writers. Overall, our goal for this course is develop students into critical thinkers and active participants in the HRI space who can develop technology meant to improve society.


\bibliographystyle{ACM-Reference-Format}
\bibliography{sample-base}


\end{document}